\documentclass[letterpaper, 10 pt, conference]{ieeeconf}  % Comment this line out if you need a4paper

\IEEEoverridecommandlockouts                              % This command is only needed if 
\usepackage{amsmath} % assumes amsmath package installed
\usepackage{amssymb}  % assumes amsmath package installed
\usepackage{cite} % For citation commands
\usepackage{algorithm}
\usepackage{algpseudocode}
\usepackage{multicol}
\usepackage{graphicx}
\usepackage{caption}
\usepackage{subcaption}
\usepackage{svg}
\usepackage{amsmath}

\title{\LARGE \bf
Real-Time Retargeting Using Controllability Boundary for Chandrayaan-3 Lunar Landing }

\author{Suraj Kumar$^{1}$, Debjyoti Chakrabarti$^{1}$, Aditya Rallapalli$^{1}$, Bharat Kumar GVP$^{1}$, Ashok Kumar Kakula$^{1}$
\thanks{$^{1}$The authors are associated with Controls and Digital Area, U R Rao Satellite Center, Indian Space Research Organization, Bengaluru, Karnataka, India \{surajk@ursc.gov.in, suraj27avionics@gmail.com\}}
}

\begin{document}

\maketitle
\thispagestyle{empty}
\pagestyle{empty}

%%%%%%%%%%%%%%%%%%%%%%%%%%%%%%%%%%%%%%%%%%%%%%%%%%%%%%%%%%%%%%%%%%%%%%%%%%%%%%%%
\begin{abstract}
This paper presents the real-time retargeting guidance policy developed for the Chandrayaan-3 lunar landing mission. The baseline guidance generates approximate fuel-optimal descent trajectories, while a high-level policy enables safe retargeting to alternate sites when the nominal site becomes infeasible. The retargeting strategy leverages a convex representation of the controllability boundary, allowing rapid feasibility checks and real-time target updates. To the best of the authors’ knowledge, this represents the first application of a data-driven retargeting framework in an operational lunar landing mission. Pre-flight simulations and Chandrayaan-3 flight results validate the effectiveness of the proposed approach. 
%This paper presents the real-time retargetting-based guidance policy developed for the Chandrayaan-3 lunar landing mission under state and actuation constraints. The baseline guidance generates fuel-efficient descent trajectories to the designated landing site, while an outer-layer policy ensures safe retargetting to alternative site if the designated site is deemed infeasible. To construct this policy, forward reachable sets are computed in a reduced state space to characterize the states that can be driven to the designated site for various perturbations in navigation and propulsion parameters. A zero-loss convex decision boundary constructed from this set enables rapid feasibility checks and target updates. To the best of authors’ knowledge, this represents the first application of a data-driven retargeting framework in an actual lunar landing mission; both simulations and Chandrayaan-3 flight results confirm its robustness, reliability, and real-time feasibility, establishing forward reachable set–based retargeting as a practical augmentation to baseline guidance for future planetary landings.
\end{abstract}

%%%%%%%%%%%%%%%%%%%%%%%%%%%%%%%%%%%%%%%%%%%%%%%%%%%%%%%%%%%%%%%%%%%%%%%%%%%%%%%%
\section{INTRODUCTION}
The Moon’s proximity to Earth makes it a strategic gateway for deep-space missions. India’s Chandrayaan-1 discovered water on the lunar surface, sparking renewed global interest. Chandrayaan-2 attempted a soft landing in 2019 but was unsuccessful. Chandrayaan-3, launched as its successor, demonstrated autonomous landing and surface mobility, successfully touching down near the lunar south pole on August $23^{rd}$, 2023, at Shiv Shakti Point located at latitude $69.373560^\circ$S and longitude $32.319750^\circ$E.

Powered Descent Guidance (PDG) technology plays a central role in planetary landing. Navigation errors and propulsion dispersions necessitate PDG policies that are both robust and adaptive. A critical capability of the guidance policy is to assess in real time whether a soft landing at the designated target is feasible and, if not, to autonomously retarget to an alternate site within the allowable landing region. 

PDG algorithms have been an active area of research for over two decades. The most prominent approaches include formulating PDG as an optimal control problem and solving them using direct methods based on convexification \cite{acikmese2007convex, sagliano2019generalized,malyuta2022convex} and indirect methods \cite{ito2020throttled,lu2023propellant,ito2023optimal,lu2024rethinking}. Analytical approaches, beginning with Apollo-era polynomial guidance \cite{klumpp1974apollo}, remain attractive for their simplicity but cannot explicitly enforce constraints; later extensions include optimal feedback laws \cite{guo2011optimal}, tunable polynomial guidance laws \cite{lu2019augmented}, learning based polynomial guidance laws \cite{kumar2026quasi}, optimal analytical terminal descent guidance laws \cite{rallapalli2025near, ito2025terminal}. The literature cited here is not meant to be exhaustive and represents only a sample of existing work on algorithmic development. In general, it may not be feasible to reach the designated landing site for all initial conditions, necessitating retargeting strategies. Convex optimization methods have been applied previously to compute constrained controllable and reachable sets \cite{eren2015constrained,srinivas2024lunar,tomita2026powered} for lunar landing. While these approaches explicitly handle constraints, their computational burden on low-speed onboard processors—already shared with navigation, control, sensing, and telemetry—limits real-time applicability.

The main contribution of this work is the development of a data-driven, real-time retargeting guidance policy based on controllable set boundary representation in a reduced state space using convex optimization. The proposed framework employs a two-level architecture: a baseline policy that generates near fuel-efficient descent trajectories, and a high-level policy that enables safe retargeting to alternate sites in real-time when the nominal site is infeasible. The algorithm is lightweight for real-time implementation and was successfully deployed during the Chandrayaan-3 lunar landing. This is, to our knowledge, the first application of a data-driven retargeting framework within the guidance layer in an operational lunar landing mission.

The paper is organized as follows: Section \ref{sec:0} provides an overview of Chandrayaan-3 descent sequence. Section \ref{sec:I} presents the problem formulation and powered descent guidance problem. Section \ref{sec:II} describes the proposed approach, covering the formulation of both the baseline guidance policy and the real-time retargeting policy. Section \ref{sec:III} presents the simulation and flight results. Section \ref{sec:IV} concludes the paper.

\section{Overview of Chandrayaan-3 Descent Sequence} \label{sec:0}
\begin{figure}[h!]
\centering
\includegraphics[width=0.45\textwidth]{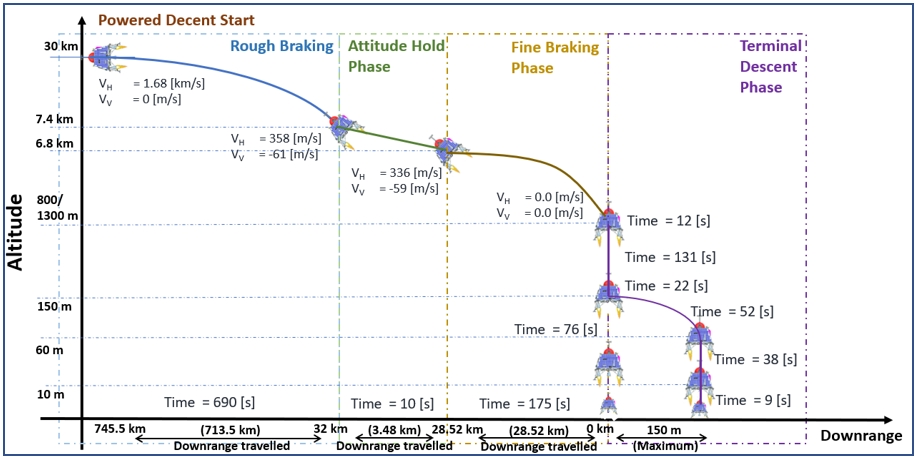}
\caption{Powered Descent phase of Chandrayaan-3}\label{pdphase}
\end{figure}
Chandrayaan-3 powered descent sequence consists of multiple phases (see Fig.~\ref{pdphase}). It begins with the Rough Braking phase from 30 km to 7.4 km altitude, where most of the initial velocity ($\sim$1.68~km/s) is reduced by firing four 800~N engines. This is followed by a 10~s Attitude Hold phase for sensor operations. The Fine Braking phase starts guiding the lander to 800~m altitude above the designated site (1.3 km altitude from the mean moon surface) with near-zero velocity. At this point, two engines are switched off and the lander hovers for 12~s, followed by vertical descent to 150~m. A second hover of up to 22~s is performed at 150~m, during which the hazard camera identifies a safe landing site. Subsequently, the lander retargets to the hazard-free site at 60~m altitude with zero velocity, then descends to 10~m with a commanded vertical velocity of $-1$~m/s. The final phase is a constant-velocity descent until touchdown, ensuring a safe and controlled landing. More details on the powered descent trajectory design can be found in \cite{kumar2026powered,rallapalli2024landing,11372411}. 

Following the attitude hold update, the navigation system is corrected using absolute sensor measurements. From this point onward, real-time decision making is required to determine whether to continue the descent to the pre-designated site or to retarget to an alternate location. The guidance development in this work is presented in the context of the Fine Braking phase, which is generally referred to as the terminal phase guidance in lunar landing literature following the main braking phase. Without focusing on mission-specific details, the discussion is kept generic for the development of a real-time retargeting policy for terminal phase guidance.
\section{Problem Formulation} \label{sec:I}
\subsection{Notations}
We adopt the following notation throughout the paper:
$\mathcal{U}$ denotes the admissible control set,
$\mathcal{X}_0$ the set of initial states,
$\mathcal{X}_f$ the hazard-free target set,
and $\mathcal{S}$ the reduced state space.
The notation $|X|$ denotes the cardinality of a set $X$,
$\hat{\boldsymbol{e}}$ denotes a unit vector,
and $\boldsymbol{x}_G$ represents the vector $\boldsymbol{x}$ expressed in the guidance frame.
\subsection{Lander Dynamics}
The lander dynamics is modeled in a local surface-fixed frame centered at the target site under the non-rotating flat-moon approximation. Since the main braking phase significantly reduces the orbital velocity, the Coriolis and lunar rotation effects can be neglected. The translational dynamics is given by
\begin{align}
\dot{\boldsymbol{r}} &= \boldsymbol{v} \nonumber \\
\dot{\boldsymbol{v}} &= \boldsymbol{g} + \boldsymbol{a} \nonumber \\
\|\boldsymbol{a}\| &= \frac{T_c}{m} \nonumber \\
\dot{m} &= -\alpha T_c
\label{lander_dynamics}
\end{align}
where $\boldsymbol{r \in \mathbb{R}^3}$, $\boldsymbol{v \in \mathbb{R}^3}$ are the position and velocity vectors relative to the target landing site; $\boldsymbol{x} = [\boldsymbol{r}^T, \boldsymbol{v}^T]^T$ is the translational state of the lander; $\boldsymbol{a} \in \mathbb{R}^3$ is acceleration due to propulsion system; $m$ is the mass of spacecraft; $\boldsymbol{g} \in \mathbb{R}^3$ is gravitational acceleration vector of the planet; $\alpha$ is positive constant associated with the fuel consumption rate; $T_c$ is the magnitude of total thrust from the propulsion system. 
\subsection{PDG Constraints}
The PDG constraints are defined by the allowable set of state and input constraints as follows: 
\begin{subequations}\label{eq:constraints}
\begin{align}
   & \rho_{m} \leq T_c \leq \rho_{M} \label{thrust_constraint}\\
   &\dot{T_c} \leq \Delta_T \label{thrust_rate} \\
   &||\boldsymbol{r}(t)|| \geq ||\boldsymbol{r}(t_f)|| \quad \forall t \in [0, t_f] \label{sub_surface} \\
   &\theta(t) \leq \theta_{lim} \label{awv} \\
   &||\boldsymbol{v} - (\boldsymbol{v}^T\boldsymbol{\hat{r}})\boldsymbol{\hat{r}}|| < v_{h,max} \label{hor_constr} \\
   &m(t_f) \geq m_T  \label{mass_constr}
\end{align}
\end{subequations}
Eq (\ref{thrust_constraint})-(\ref{thrust_rate}) is associated with the physical limits imposed by the propulsion system. $\rho_{m}, \rho_{M}$ denote the minimum and maximum thrust of the engines, and $\Delta_T$ denotes the upper bound on the rate of change in thrust. Eq (\ref{sub_surface}) requires the lander trajectory not to go below the target to avoid surface collision. Eq (\ref{awv}) is the attitude constraint that limits the deviation in the thrust vector from the vertical direction to ensure the proper operation of the line-of-sight sensors. Eq (\ref{hor_constr}) limits the maximum horizontal velocity for sensor operations. Finally, Eq (\ref{mass_constr}) requires that the terminal mass at the end of the guidance phase be upper bounded by a threshold ($m_T$). The boundary constraints are given as   
\begin{subequations}\label{boundary_conditions}
\begin{align}
    & \boldsymbol{x}(t_0) \in \mathcal{X}_0, \quad \boldsymbol{x}(t_f) \in \mathcal{X}_f \label{pos_cnstr}\\
    & \boldsymbol{a}(t_0) = \boldsymbol{a}_0, \quad \boldsymbol{a}(t_f) = \boldsymbol{a}_f \label{acc_cnstr}
\end{align}
\end{subequations}
Nominally, the lander is designated to reach a single target $\boldsymbol{x}^*_f \in \mathcal{X}_f$. The set $\mathcal{X}_0$ is modeled by a Gaussian process with mean given by the nominal phase-start vector and covariance determined by inertial navigation drift during the prime braking phase and orbit determination errors. Boundary constraints on acceleration are imposed to ensure continuity in acceleration across the phase.
\subsection{Powered Descent Guidance Problem}
The powered descent guidance (PDG) objective is to compute a feasible control input $\boldsymbol{a}(t) \in \mathcal{U}$ that drives the lander from an initial condition $\boldsymbol{x}(0) \in \mathcal{X}_0$ to a state as close as possible to the designated site, while satisfying dynamics (\ref{lander_dynamics}), constraints (\ref{eq:constraints}) and boundary conditions (\ref{boundary_conditions}). 

When the objective is to maximize terminal mass, the problem is referred to as \emph{Fuel-Optimal Powered Descent Guidance (FOPDG)}, which is posed as:
\begin{equation}
\begin{aligned}
& \min_{\boldsymbol{a}(\cdot) \in \mathcal{U}} \; -m(t_f) \\
& \text{subject to:} \\
& \quad \text{Dynamics (\ref{lander_dynamics}) with } \boldsymbol{x}(0) \in \mathcal{X}_0, \\
& \quad \text{Constraints (\ref{eq:constraints}) and (\ref{boundary_conditions})}
\end{aligned}
\label{eq:pdg_problem}
\end{equation}
Our goal is to design a tractable analytical guidance law that approximates the solution of (\ref{eq:pdg_problem}). To this end, we decompose the PDG problem into two subproblems:

\textbf{Problem 1 (Base Guidance Policy)}: For a designated target $\boldsymbol{x}^*_f \in \mathcal{X}_f$, design an approximate solution to (\ref{eq:pdg_problem}). Let $\pi_{\text{b}}$ denote the base guidance policy. This policy generates control actions $\boldsymbol{a}(t) = \pi_{\text{b}}(\boldsymbol{x}(t), t)$. 

\textbf{Problem 2 (Real Time Retargeting Policy)}: Design a policy that, given $\boldsymbol{x}(0) \in \mathcal{X}_0$, assesses the feasibility of reaching $\boldsymbol{x}_f^*$ using the controllable set spanned by the space of the base policy. If infeasible, identify an alternative reachable site $\boldsymbol{\tilde{x}}_f \in \mathcal{X}_f$ in real time by searching for the reachable set of $\boldsymbol{x}(0)$. 
\subsection{Constrained Controllable and Reachable Sets}
\textbf{Definition-1 (Constrained Reachable Set).}  
Given an initial state $\boldsymbol{x}(t_0)$ at time $t_0$, the Constrained Reachable Set (CRS) under the action of the base policy is defined as
\begin{equation}
\begin{aligned}
\mathcal{X}_R^{\pi_{b}}(t_f; \boldsymbol{x}(t_0)) = \big\{ \boldsymbol{x}(t_f) \;\big|\; & \exists \, \boldsymbol{a}(\cdot) \in \pi_{b} \;\; \text{s.t.} \\
& \boldsymbol{x}(\cdot) \text{ satisfies } (\ref{lander_dynamics}), (\ref{eq:constraints}), (\ref{boundary_conditions}) \big\}
\end{aligned}
\end{equation}
That is, $\mathcal{X}_R^{\pi_{b}}$ consists of all admissible states that can be reached from the initial state $\boldsymbol{x}(t_0)$ at the terminal time $t_f$.
\textbf{Definition-2 (Constrained Controllable Set).}  
For a desired target state $\boldsymbol{x}_f^*$ at the terminal time $t_f$, the Constrained Controllable Set (CCS) is defined as
\begin{equation}
\begin{aligned}
\mathcal{X}_C^{\pi_{b}}(t_f; \boldsymbol{x}_f^*) = \big\{ \boldsymbol{x}(t_0) \;\big|\; & \exists \, \boldsymbol{a}(\cdot) \in \pi_b \;\; \text{s.t.}
\boldsymbol{x}(t_f) = \boldsymbol{x}_f^*, \;\; \\ 
& \boldsymbol{x}(\cdot) \text{ satisfies } (\ref{lander_dynamics}), (\ref{eq:constraints}), (\ref{boundary_conditions}) \big\}
\end{aligned}
\end{equation}
That is, $\mathcal{X}_C^{\pi_{b}}$ is the set of all admissible initial states $\boldsymbol{x}(t_0)$ that can be exactly steered to the prescribed target at the terminal time $t_f$.

The real-time retargeting policy (Problem 2) acts as an outer layer on the base guidance policy (Problem 1), which can be implemented using any tractable method. If the desired target $\boldsymbol{x}_f^*$ is infeasible, i.e., $\boldsymbol{x}(t_0) \notin \mathcal{X}_C^{\pi_b}(t_f;\boldsymbol{x}_f^*)$, then a new feasible target $\tilde{\boldsymbol{x}}_f$ is selected by projecting $\boldsymbol{x}(t_0)$ onto the controllability boundary $\partial \mathcal{X}_C^{\pi_b}$. This ensures that the perturbed initial state lies within the controllable set of $\tilde{\boldsymbol{x}}_f$, allowing continued execution of the base policy without recomputation. This modular structure makes the retargeting framework generic.

\section{Proposed Approach}
\label{sec:II}
\subsection{Base Guidance Policy}

The guidance policy that solves Problem~1 has been extensively studied in the literature, with a focus on both iterative and non-iterative solutions. For the development of the real-time retargeting policy, any suitable guidance scheme can be employed. Our first contribution is the analytical guidance policy developed for Chandrayaan-3, which significantly extends the robustness and fuel efficiency of the classical polynomial guidance through simulation-driven optimization.

We first address the feasibility problem of determining a feedback guidance policy that satisfies the system dynamics and boundary conditions. This is achieved by prescribing a sufficiently high-order polynomial and computing its coefficients to meet these constraints. Within the resulting family of feasible solutions, fuel efficiency is achieved by introducing a free parameter for optimization. In the current formulation, this free parameter is the time-to-go ($t_{go} = t_f - t$), representing the remaining time to reach the target.

Without loss of generality, the 3D guidance problem can be projected onto a 2D plane defined by the current position vector $\boldsymbol{r}_0$ and target position vector $\boldsymbol{r}_f$ that defines the motion direction. Any residual velocity component orthogonal to this plane is corrected using a regulation control strategy. 
Given two non-parallel vectors \( \boldsymbol{r}_0 \) and \( \boldsymbol{r}_f \), we construct an orthonormal basis as follows:
\begin{equation}
\hat{\mathbf e}_x=\frac{\mathbf r_0}{\|\mathbf r_0\|}, \quad
\hat{\mathbf e}_z=\frac{\hat{\mathbf e}_x\times\mathbf r_f}{\|\hat{\mathbf e}_x\times\mathbf r_f\|}, \quad
\hat{\mathbf e}_y=\hat{\mathbf e}_z\times\hat{\mathbf e}_x
\end{equation}
%\begin{equation}
%\begin{aligned}
%    \boldsymbol{\hat{e}_x} &= \frac{\boldsymbol{r}_0}{\|\boldsymbol{r}_0\|}, \quad
%    \boldsymbol{\hat{e}_y} = \frac{\boldsymbol{r}_f}{\|\boldsymbol{r}_f\|}, \\
%    \boldsymbol{\hat{e}_z} &= \frac{\boldsymbol{\hat{e}_x} \times \boldsymbol{\hat{e}_y}}{\|\boldsymbol{\hat{e}_x} \times \boldsymbol{\hat{e}_y}\|}, \quad
%    \boldsymbol{\hat{e}_x} = \frac{\boldsymbol{\hat{e}_y} \times \boldsymbol{\hat{e}_z}}{\|\boldsymbol{\hat{e}_y} \times \boldsymbol{\hat{e}_z}\|}
%\end{aligned}
%\end{equation}

The transformation matrix \( \mathbf{T} \in \mathbb{R}^{3 \times 3} \) used to transform a vector from the navigation frame to the guidance frame is then given by:
\begin{equation}
\mathbf{T_G} = \begin{bmatrix}
\boldsymbol{\hat{e}_x} & \boldsymbol{\hat{e}_y} & \boldsymbol{\hat{e}_z}
\end{bmatrix}
\end{equation}
Considering the boundary constraints (\ref{boundary_conditions}) in position, velocity and acceleration appropriately transformed in guidance, we prescribe a cubic polynomial for the net acceleration vector in the guidance frame, 
\begin{equation}
    \boldsymbol{a_G}(t) = \boldsymbol{C}_{0} + \boldsymbol{C}_{1}t + \boldsymbol{C}_{2}t^2 + \boldsymbol{C}_{3}t^3
\end{equation}
We then solve for the feasible solution by enforcing dynamics and boundary conditions through a series of linear systems of equations
\begin{equation}
\label{gui_lin_eq}
    \begin{bmatrix} 
    1 & 0 & 0 & 0 \\
    1 & t_{\text{go}} & t_{\text{go}}^2 & t_{\text{go}}^3 \\
    t_{\text{go}} & \frac{t_{\text{go}}^2}{2} & \frac{t_{\text{go}}^3}{3} & \frac{t_{\text{go}}^4}{4} \\
    \frac{t_{\text{go}}^2}{2} & \frac{t_{\text{go}}^3}{6} & \frac{t_{\text{go}}^4}{12} & \frac{t_{\text{go}}^5}{20} \\
\end{bmatrix}\begin{bmatrix} C_{0}^i \\ C_{1}^i \\ C_{2}^i \\ C_{3}^i \end{bmatrix} = \begin{bmatrix}
\boldsymbol{a}_{G,0}^i \\
\boldsymbol{a}_{G,f}^i \\
\boldsymbol{v}_{G,f}^i - \boldsymbol{v}_{G,0}^i \\
\boldsymbol{r}_{G,f}^i - \boldsymbol{r}_{G,0}^i - \boldsymbol{v}_{G,0}^i \cdot t_{go} \\
\end{bmatrix} \\
\end{equation}
where $t_{go} = t_f - t$ denotes the remaining time in the powered descent.
Index $i$ represents the indices of the $x$, $y$, $z$ components of the vector.
In each guidance cycle, the net acceleration is calculated by substituting the current states of the lander appropriately transformed in the guidance frame as the initial states to determine the polynomial coefficients and $t_{go}$ is linearly decremented by each guidance cycle. 

The policy for the free parameter $t_{go}$ is obtained using simulation-driven optimization in a supervised learning framework. Specifically, the $t_{go}$ policy is defined as a mapping from the guidance state represented as downrange ($S$), altitude ($H$), vertical velocity ($w$), and horizontal velocity ($v$), to an optimal value of $t_{go}$ that consumes minimum fuel. These four parameters are readily obtained from the guidance state $(\boldsymbol{r}_G, \boldsymbol{v}_G)$ and uniquely characterize the motion of the lander in the projected 2D plane. Formally, 
\begin{equation}
    t_{go} = \pi_{tgo}(S, H, w, v)
\end{equation}
where $\pi_{tgo}(\cdot)$ denotes the learned policy.

We now describe the pipeline for constructing the supervised learning dataset. As will be evident in the subsequent discussion, this procedure not only generates the training data but also yields, the characterization of the controllable set under the action of the base guidance policy.

Let $\boldsymbol{\bar{x}}_G$ be the nominal guidance state vector at the start of the guidance phase and $\Sigma_G$ the covariance matrix representing the dispersions in the guidance states. Since our objective is to model the relation between $t_{go}$ and the guidance states, we construct the guidance dataset $\Xi$ as
\begin{equation}
\begin{aligned}
    \Xi := \{\boldsymbol{\bar{x}}_G + \delta_i \;|\; \delta_i \sim \mathcal{N}(0, \Sigma_G), \; i = 1,2,\ldots,|\Xi| \}
\end{aligned}
\end{equation}
For each perturbed state $x_{G,i} \in \Xi$, the cubic polynomial guidance is rolled out over a specified range of $t_{go}$ between $t^{min}_{go}$ and $t^{max}_{go}$ to determine whether the lander converges to the desired terminal conditions under the prescribed state and action constraints (Eq.~\ref{eq:constraints}). This yields the set $\mathcal{F}(x_{G,i})$ as
\begin{equation}
\begin{aligned}
    \mathcal{F}(\boldsymbol{x}_{G,i}) = \{(t_{go,k}, m_{f,k}) \;|\; k = 1,2,\ldots,|\mathcal{F}(\boldsymbol{x}_{G,i}|\}
\end{aligned}
\end{equation}
where $m_f$ denotes the final mass in that trajectory rollout. Each element of this set represents the tuple of $t_{go}$ and terminal mass for the given state vector. Algorithm \ref{algo:1} details the procedure of generating this set.   
The optimal $t_{go}^*(x_{G,i})$, corresponding to the fuel-optimal trajectory for state $x_{G,i}$, is then obtained as
\begin{equation}
\label{opt_tgo}
    t_{go}^{*}(x_{G,i}) = \arg\max_{(t_{go}, m_f) \in \mathcal{F}(x_{G,i})} m_f
\end{equation}
The supervised learning dataset $\mathcal{D}$ is then given as
\begin{equation}
\begin{aligned}
    \mathcal{D} := \{(x_{G,j}, \; t_{go}^*(x_{G,j})) \;|\; \; j=1,2,\ldots,|\Xi|\}.
\end{aligned}
\end{equation}
Finally, the constrained controllable set is given by the union of all nonempty sets $\mathcal{F}(\boldsymbol{x}_{G,i})$:
\begin{equation}
\label{cont_set}
    \mathcal{X}_C^{\pi_b} := \bigcup_{x_{G,i} \in \Xi} \mathcal{F}(x_{G,i})
\end{equation}
Formally, $\pi_{tgo}$ policy is expressed as:
\begin{equation}
\label{general_tgo_eqn}
    \pi_{tgo}(\boldsymbol{x}_{G_i}) = K^T\phi(H,S,w,v; \mathcal{D})
\end{equation}
where $\phi$ defines the function approximator used for fitting the data. In general, a polynomial function approximator is sufficient, though any suitable function approximator, including neural networks, can be employed. For Chandrayaan-3, a second-order polynomial was adequate. Note that for evaluation of the above expression $H, S, w \text{ and } v$ are first obtained from the guidance state. The coefficient vector K is arrived at using lasso regression on the dataset $\mathcal{D}$ as
\begin{equation}
\min_{K} \left( \frac{1}{2} \sum_{i=1}^{|\mathcal{D}|} \left( t_{\text{go},i} - \pi_{tgo}(\boldsymbol{x}_{G,i}) \right)^2 + \mu \|K\|_1 \right)
\end{equation}
where $\mu \|K\|_1$  is the regularization term, $\|K\|_1$ denotes the $l_1$-norm of the coefficient vector K and $\mu$ governs the trade-off between model fidelity and complexity. In particular, it controls the strength of the $\ell_1$-norm penalty applied to the regression coefficients, promotes sparsity in the resulting model, making it suitable for on-board implementation. The cubic polynomial, together with the offline optimized $t_{go}$ policy, defines the base guidance policy.

In essence, this approach approximates the solution to the FOPDG problem described in Eq (\ref{eq:pdg_problem}) by restricting the acceleration to a class of functions parameterized by polynomials. Naturally, it is less optimal than the globally fuel-optimal solution, since optimization identifies a fuel-efficient trajectory only within the restricted class of cubic polynomials.
\begin{algorithm}[h]
\caption{Algorithm for $\mathcal{F}$ set computation}
\label{algo:1}
\begin{algorithmic}[1]
\State \textbf{Input:}$\boldsymbol{x_G}$, $t_{go}^{min}$, $t_{go}^{max}$, hyperparameters $\epsilon_T$, $\delta t_{go}$
\State \textbf{Output:} $\mathcal{F}(\boldsymbol{x_G})$
    \State Define convergence indicator function for trajectory rollout: 
\Statex $\mathcal{I}_{\text{c}} =
\begin{cases}
1, & \text{if} \quad
\begin{aligned}
\text{Constraints} (\ref{eq:constraints}), (\ref{boundary_conditions}) \text{ satisfied}
\end{aligned}
\\
0, & \text{otherwise}
\end{cases}
$
\State \textbf{Initialize:} $\mathcal{F}(S) \gets \emptyset$, $t_{go,i}^{min} \gets t_{go}^{min}$, $t_{go,i}^{max} \gets t_{go}^{max}$ 
    %\State Bisection phase:
    \While{$(t_{go,i}^{max}- t_{go,i}^{min}) > \epsilon_T$} %\Comment{Bisection search for computation of minimal $t_{go}$}
        \State $t_{go}^{cur} \gets \frac{t_{go,i}^{min} +\, t_{go,i}^{max}}{2}$
        \State Perform trajectory rollout with $t_{go}^{cur}$
        \State Obtain terminal mass $m_f$ and evaluate $\mathcal{I}_c$ 
        \If{$\mathcal{I}_{\text{c}}==1$}
            \State  $t_{go,i}^{max} \gets t_{go}^{cur}$
        \Else
            \State $t_{go,i}^{min} \gets t_{go}^{cur}$
        \EndIf
    \EndWhile
    \State Record $(t_{go,i}^{min}, m_{f})$ in $\mathcal{F}(\boldsymbol{x}_G)$
    \State Initialize: $t_{go,i}^{max} \gets t_{go}^{max}$ 
    \While{$(t_{go,i}^{max}-t_{go,i}^{min}) > \epsilon_T$} %\Comment{Construction of set $\mathcal{F}$}
         \State Perform trajectory rollout with $t_{go,i}^{max}$
    \State Obtain terminal mass $m_{f,i}$ and evaluate $\mathcal{I}_c$
        \If{$\mathcal{I}_{\text{c}}==1$}
            \State Record $(t_{go,i}^{max}, m_{f,i})$ in $\mathcal{F}(\boldsymbol{x}_G)$ 
        \EndIf
        \State Update $t_{go,i}^{max} \gets t_{go,i}^{max}-\delta t_{go}$
    \EndWhile

\State $\Return \, \mathcal{F}(\boldsymbol{x}_G)$
\end{algorithmic}
\end{algorithm}
\subsection{Real-Time Retargeting Policy}
The key idea of the proposed approach is to represent the controllability boundary in a reduced state space using a convex function. This function separates the controllable set from the uncontrollable region and serves as a decision surface for assessing the feasibility of landing at the designated site. When the designated target $\boldsymbol{x}_f^*$ is infeasible, i.e., $\boldsymbol{x}(t_0) \notin \mathcal{X}_C^{\pi_b}(t_f;\boldsymbol{x}_f^*)$, a new feasible target $\tilde{\boldsymbol{x}}_f$ is obtained by projecting the perturbed initial state $\boldsymbol{x}(t_0)$ onto the controllability boundary $\partial \mathcal{X}_C^{\pi_b}$. Unlike approaches that explicitly construct the reachable set of $\boldsymbol{x}(t_0)$, which is computationally intensive online, this projection guarantees at least one feasible solution lying inside the reachable set. While multiple feasible targets may exist with different fuel penalty, the proposed method guarantees that the system remains within the controllable domain of the updated target while allowing continued execution of the base policy without recomputation or excessive fuel consumption. The main advantage of this framework is that it eliminates the need to recompute the base policy in real time, which is particularly important when the base guidance law is derived through iterative or computationally intensive methods. Therefore, real-time retargeting policy design involves two steps: a zero-loss convex decision boundary that parametrizes the controllability boundary and a projection operator. 

Since the admissible set of initial conditions is assumed to follow a Gaussian process, the controllability boundary is parameterized using a convex conic function. Owing to the reduced state space representation, the controllability dataset is generally not linearly separable, even though it admits tractable analytical solutions. Consequently, a bi-level optimization strategy is employed to handle this nonlinearity. In the first level, a maximum-margin classifier is employed to identify a separating decision boundary in the reduced state space, while in the second level, the boundary is shifted to eliminate all misclassifications, thereby yielding a zero-loss convex boundary.

\textbf{Level-1 Optimization:} 
We first transform the controllable set (Eq.~\ref{cont_set}) into a supervised binary classification problem, where each state sample is labeled according to whether it lies inside or outside the controllability boundary. Formally, the dataset $\mathcal{T}$ for classification is defined as
\begin{equation}
\label{eq:svm_dataset}
\mathcal{T} = \{ (\boldsymbol{x}_i, d_i) \mid \boldsymbol{x}_i \in \mathcal{X}_0,\; d_i \in \{-1, +1\} \}
\end{equation}
where $d_i$ is the class label, with $d_i = +1$ denoting controllable states and $d_i = -1$ denoting uncontrollable states. We consider a conic function to represent the convex decision boundary in a reduced-dimensional state space. Let the reduced state vector be defined as
\[
\mathbf{s} = \begin{bmatrix} s_1 \\ s_2 \end{bmatrix} =
\begin{bmatrix} S/v \\ H/w \end{bmatrix}
\]
where $s_1$ and $s_2$ denote the ratio of range to horizontal velocity and altitude to vertical velocity, respectively. This reduction preserves the dominant lander kinematic features that determine controllability. The convex decision boundary is expressed as
\begin{align} \label{decision_boundary}
    f(c,b;\,Z) = c^{T}Z + b
\end{align}
where the feature vector $Z$ for the $i^{\text{th}}$ sample is given by
\begin{align} \label{conic}
Z_i = \begin{bmatrix}
    s_{1,i} & s_{2,i} & s_{1,i}^2 & s_{2,i}^2 & s_{1,i}s_{2,i}
\end{bmatrix}^T, \quad i = 1, \ldots, n.
\end{align}
The coefficients $c$ and $b$ are data-dependent parameters obtained from the labeled dataset $\mathcal{T}$. The objective is to identify a separating boundary that maximizes the geometric margin between controllable and uncontrollable states. This is posed as the maximum-margin classification problem\cite{cristianini2000introduction}:
\begin{eqnarray}
&\underset{c,b}{\text{Minimize}} &\frac{1}{2}\|c\|^2 + \gamma \sum\limits_{i=1}^n \zeta_i \nonumber\\
&\text{Subject to} & d_i(c^{T}Z_i+b) \geq 1 - \zeta_i, \quad i = 1, \ldots, n \nonumber \\
& & \zeta_i \geq 0, \quad i = 1, \ldots, n \label{svm}
\end{eqnarray}
where $\gamma$ is a regularization parameter that controls the trade-off between maximizing the margin and penalizing classification errors; $\zeta_i$ is the slack variable associated with misclassification of $i^{th}$ data.

\textbf{Level-2 Optimization:}
From the perspective of data classification, the Level-1 optimization provides an optimal separating surface by maximizing the margin between controllable and uncontrollable datasets. However, for planetary landing applications, even a single false classification may lead to mission failure. Consequently, a \textit{zero-tolerance policy} toward misclassification is adopted. Simply tuning the hyperparameter $\gamma$ in Eq~(\ref{svm}) cannot guarantee this requirement, since arbitrary shifts of the decision boundary within the maximum-margin region may misclassify controllable states as uncontrollable, thereby unnecessarily shrinking the operating region of the base guidance policy. Therefore, a second optimization stage is introduced to adjust the decision boundary such that misclassification is eliminated while minimizing shrinkage of the controllable region.

The decision boundary obtained from the Level-1 optimization in Eq~(\ref{svm}) defines a conic surface separating controllable and uncontrollable states in the reduced state space. To facilitate subsequent manipulation of this boundary, we rewrite it in the form of a general conic:
\begin{equation}
    As_1^2 + Bs_1s_2 + Cs_2^2 + Ds_1 + Es_2 + F = 0
    \label{gen_conic}
\end{equation}
where the coefficients \(A,B,C,D,E,F\) are obtained from \((c,b)\).
To obtain a more convenient representation, the conic is first translated to its centroid. Consider the coordinate transformation
\begin{equation}
    s_1 = \bar{s}_1 + h, \qquad s_2 = \bar{s}_2 + k
    \label{origin_transform}
\end{equation}
where $(h,k)$ denotes the centroid of the conic. Substituting Eq~(\ref{origin_transform}) into Eq~(\ref{gen_conic}) yields
\begin{equation}
    \bar{A} \bar{s}_1^2 + \bar{B} \bar{s}_1 \bar{s}_2 + \bar{C} \bar{s}_2^2 = 1
    \label{translated}
\end{equation}
where $\bar{A}=-A/\lambda$, $\bar{B}=-B/\lambda$ and $\bar{C}=-C/\lambda$. The parameters $\lambda$, $h$, and $k$ are given by
\begin{align}
\lambda &= Ah^2 + Bhk + Ck^2 + Dh + Ek + F \\
h &= \frac{2CD - BE}{B^2 - 4AC}, \qquad
k = \frac{2AE - BD}{B^2 - 4AC}
\label{centroid}
\end{align}
The translated conic in Eq~(\ref{translated}) still contains the cross-term $\bar{s}_1\bar{s}_2$. This term can be eliminated by rotating the coordinate system to align with the principal axes of the conic. The rotation is defined as
\begin{equation}
\begin{bmatrix}
\bar{s}_1\\
\bar{s}_2
\end{bmatrix}
=
R
\begin{bmatrix}
x\\
y
\end{bmatrix}, \qquad
R=
\begin{bmatrix}
\cos\theta & \sin\theta\\
-\sin\theta & \cos\theta
\end{bmatrix}
\label{rot1}
\end{equation}
where the rotation angle $\theta$ satisfies
\begin{equation}
\tan\theta=
\frac{(\bar{C}-\bar{A})+\sqrt{\bar{B}^2+(\bar{C}-\bar{A})^2}}
{\bar{B}}
\label{rot_angle_conic}
\end{equation}
Substituting Eq~(\ref{rot1}) into Eq~(\ref{translated}) yields the principal-axis representation
\begin{equation}
    \bar{s}_1^2 + \bar{s}_2^2 = 1
    \label{standard_conic}
\end{equation}
which corresponds to a standard conic form of the decision boundary. Rewriting Eq~(\ref{standard_conic}) in terms of the original coordinates gives
\begin{align}
\label{standard_conic2}
M_1\big(&\cos\theta(x-h)+\sin\theta(y-k)\big)^2 \notag\\
&+M_2\big(-\sin\theta(x-h)+\cos\theta(y-k)\big)^2=1 \\
M_1 &= A_1\cos^2\theta + B_1\cos\theta\sin\theta + C_1\sin^2\theta \\
M_2 &= A_1\cos^2\theta - B_1\cos\theta\sin\theta + C_1\sin^2\theta 
\end{align}
Using Eq~(\ref{standard_conic2}), the Level-1 decision function corresponding to Eq~(\ref{decision_boundary}) can be written as
\begin{align}
f(c,b;Z)=&\;
M_1\big(\cos\theta(x-h)+\sin\theta(y-k)\big)^2 \notag\\
&+M_2\big(-\sin\theta(x-h)+\cos\theta(y-k)\big)^2-1
\label{svm2}
\end{align}
The zero-level set of $f(c,b;Z)$ defines the maximum-margin separation between the controllable and uncontrollable sets. In this representation, $(h,k)$ denote the centroid of the conic and $\theta$ its principal orientation.  Instead of directly modifying the coefficients $(c,b)$, the boundary is parameterized geometrically through its centroid and orientation. A small perturbation $\delta=[\delta_h,\delta_k,\delta_\theta]^T$ produces a perturbed decision function $g(\delta;Z)$, which is equivalent to evaluating $f(c,b;Z)$ under the corresponding geometric perturbation. The perturbed decision boundary is given by
\begin{align}
g(\delta;Z)=&\;
M_1(\delta)\big(\cos\theta'(x-h')+\sin\theta'(y-k')\big)^2 \notag\\
&+M_2(\delta)\big(-\sin\theta'(x-h')+\cos\theta'(y-k')\big)^2-1
\label{svm3}
\end{align}
where $h'=h+\delta_h$, $k'=k+\delta_k$, and $\theta'=\theta+\delta_\theta$. The function $g(\delta;Z)$ therefore represents a perturbed version of the Level-1 decision boundary used to enforce the zero-misclassification constraint.

The optimal perturbation for zero-misclassification is determined through the following Level-2 optimization problem
\begin{eqnarray}
\min_{\delta} && \delta^T\delta \nonumber\\
\text{s.t.} && g(\delta;Z^{(i)})\le \eta_i,\quad i=1,\ldots,n
\label{shift_optim}
\end{eqnarray}
where $\eta \in \mathbb{R}^n$ serves as a hyperparameter that balances the trade-off between minimizing misclassification loss and preserving the size of the controllable set. The objective seeks the smallest possible perturbation of the Level-1 boundary while satisfying the classification constraints imposed by the dataset.

Solving Eq~(\ref{shift_optim}) yields the optimal perturbation $\delta^\star$, resulting in the zero-loss decision boundary
\begin{equation}
g(\delta^\star;Z)=0
\label{svm4}
\end{equation}
Accordingly, the controllability boundary under the base guidance policy $\pi_b$ is given as
\begin{equation}
\partial\mathcal{X}_C^{\pi_b}(t_f;\boldsymbol{x}_f^*)=
\{z\in\mathcal{S}\mid g(\delta^\star;z)=0\}
\label{eq:cboundary}
\end{equation}
and the corresponding controllable set is
\begin{equation}
\mathcal{X}_C^{\pi_b}(t_f;\boldsymbol{x}_f^*)=
\{z\in\mathcal{S}\mid g(\delta^\star;z)>0\}
\label{eq:inside_outside}
\end{equation}

\textbf{Real-time retargeting:}
Once the controllability boundary is obtained, the next step is to determine the required retargeting when the current state lies outside the controllable set. Since only the range component must be modified, the retargeting reduces to a horizontal projection
\begin{equation}
\mathbb{P}(s_1,s_2)=(s_1',s_2)
\label{projection}
\end{equation}
To minimize deviation from the nominal landing target while guaranteeing convergence, the projection is performed onto the controllability boundary itself. Because the Level-2 optimization ensures that this boundary lies strictly within the convergence region of the base guidance policy, any point projected onto it is guaranteed to converge.

All computationally intensive steps, including dataset generation, SVM training, and boundary optimization, are performed offline during mission design. The resulting controllability boundary is stored onboard in analytical form through the conic decision function $g(\delta^\star;z)$. During flight, retargeting requires only evaluating this function and solving a quadratic equation to obtain the projected range coordinate while keeping altitude and velocity states fixed. Consequently, the onboard computation involves only a small number of arithmetic operations and is compatible with the limited processing capability of typical spacecraft flight computers.

\section{Simulation and Flight Results}
\label{sec:III}
Extensive simulation experiments have been carried out to verify and validate the performance of the proposed guidance algorithm. %Fig \ref{gui_arch} shows the control flow diagram of the proposed guidance scheme. The parameters associated with the problem are the following.
\begin{equation}
\begin{aligned}
\mathbf{g} &= \begin{bmatrix} -1.68 & 0 & 0 \end{bmatrix}^T \ \text{m/s}^2, \quad m_{\text{wet}} = 1050 \ \text{kg}, \\
\alpha &= 0.00035 , \quad T_{min} = 1480 \text{N}, \quad T_{max} = 3120 \text{N} \\
\end{aligned}
\label{eq:simparams}
\end{equation}

%\begin{figure}[h]
%\centering
%\includegraphics[width=0.4\textwidth]{gui_arch.drawio.pdf}
%\caption{Proposed Guidance Control flow diagram}\label{gui_arch}
%\end{figure}

The guidance nominal state, covariance parameters and dispersion bounds for controllable set generation are given in Table \ref{tab:gui_inp_data}. Note that although the covariance in the nominal guidance state is small (third column), a larger dispersion bound is used in simulation to capture the full controllable set under the base policy. This is, in general, a standard practice in quantifying the limit of performance of the algorithm for landing applications. 
\begin{table}[hbt!]
\caption{\label{tab:gui_inp_data} Guidance Parameters}
\centering
\begin{tabular}{lcccccc}
\hline
Parameter & Nominal & $3\sigma$ Error & Dispersion bound \\\hline
Altitude [km] & 6.8 & $\pm0.9$ & $\pm3.0$ \\
Range [km] & 28.5 & $\pm0.75$ & $\pm3.0$ \\
Vertical Velocity [m/s] & 59 & $\pm2.5$ & $\pm17$ \\
Horizontal Velocity [m/s] & 336 & $\pm1.75$ & $\pm17$ \\
\hline
\end{tabular}
\end{table}
\begin{figure}[h!]
\centering
\includegraphics[width=0.35\textwidth]{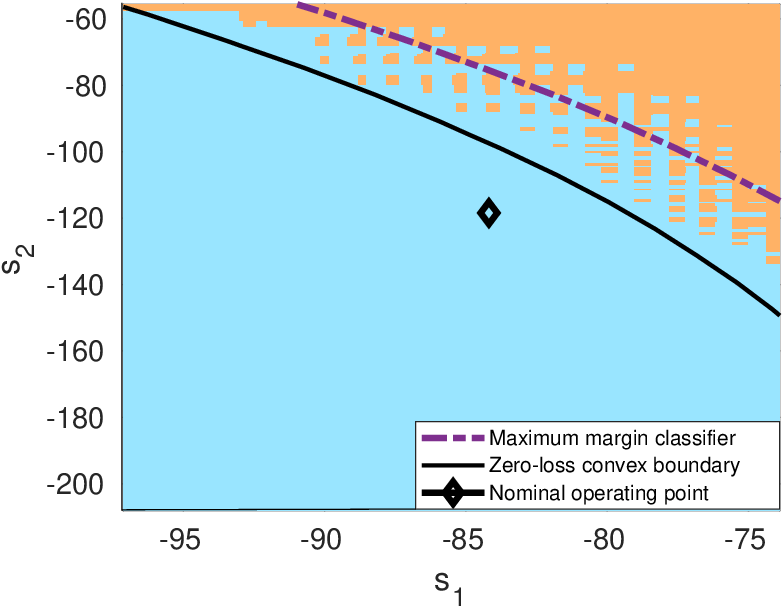}
\caption{Controllability Boundary}\label{decision_boundary3}
\end{figure}

Fig.~\ref{decision_boundary3} illustrates the controllable set and its boundary in the reduced state space, generated using Table~\ref{tab:gui_inp_data} and Algorithm~\ref{algo:1}. The nominal operating point is indicated by a diamond marker. The blue region denotes the controllable set, while the orange region corresponds to the uncontrollable set. As expected, the data is not linearly separable due to dimensionality reduction. The magenta curve represents the maximum-margin classifier, and the black curve denotes the zero-loss convex boundary.
\begin{figure}[t!] 
\centering 
    \begin{minipage}{.22\textwidth} 
    \centering 
    \includegraphics[width=0.94\textwidth]{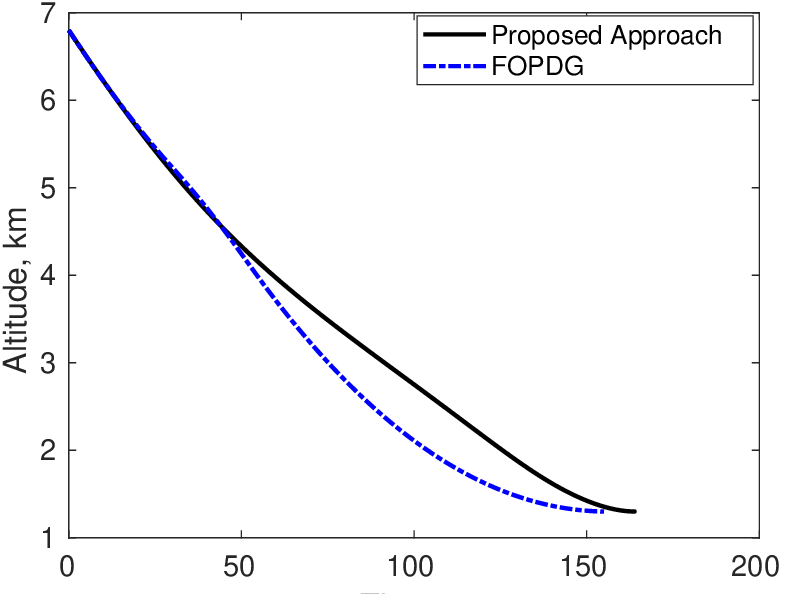} 
    \subcaption{Altitude}
    \label{fig:rtr_alt} 
    \end{minipage} 
    \begin{minipage}{.22\textwidth} 
    \centering 
    \includegraphics[width=1\textwidth]{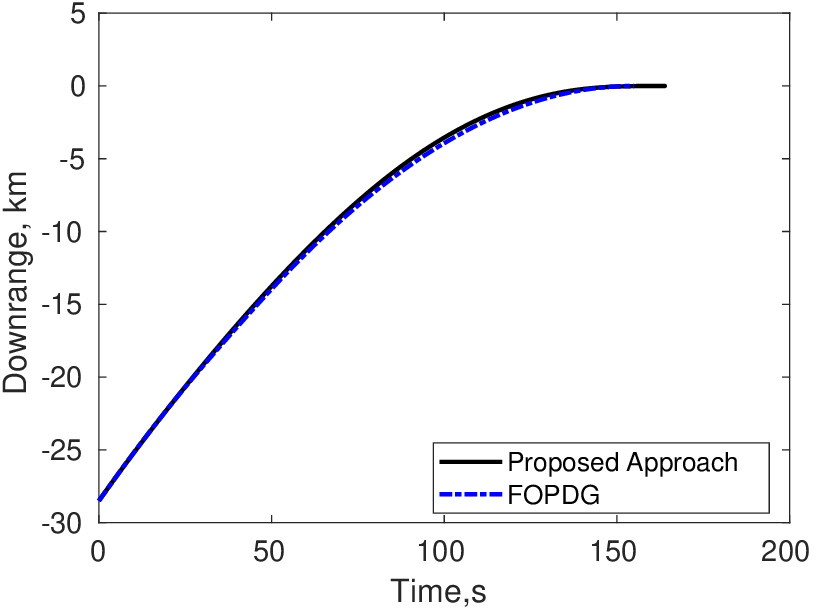} 
    \subcaption{Downrange}
    \label{fig:rtr_thr} 
    \end{minipage} 
    \begin{minipage}{.22\textwidth} 
    \centering 
    \includegraphics[width=1\textwidth]{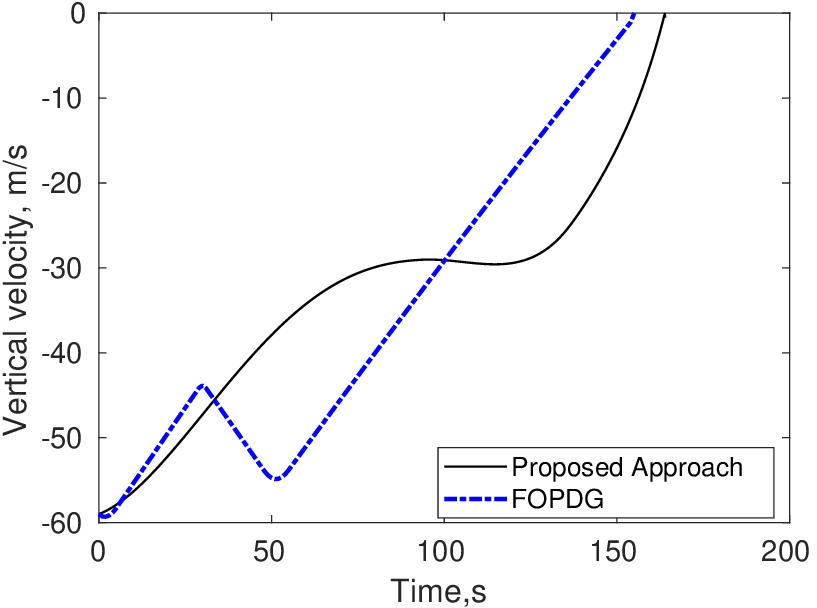} 
    \subcaption{Vertical Velocity}
    \label{fig:rtr_vv} 
    \end{minipage} 
    \begin{minipage}{.22\textwidth} 
    \centering 
    \includegraphics[width=1\textwidth]{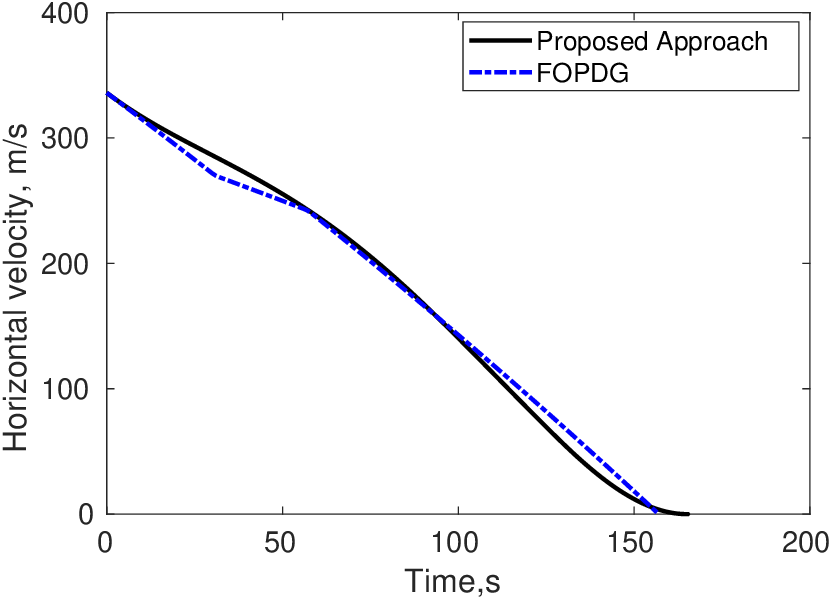} 
    \subcaption{Horizontal Velocity}
    \label{fig:rtr_hv} 
    \end{minipage} 
\caption{Comparison of trajectory traces for the Proposed Guidance and Fuel-Optimal Powered Descent Guidance with simulation parameters: 
$\mathbf{r_{G,0}}=[-28.5,0,6.8]^T$ km, 
$\mathbf{v_{G,0}}=[336,0,-59]^T$ m/s, 
$\mathbf{a_{G,0}}=[-2.26,0.0,1.91]^T$ m/s$^2$; 
$\mathbf{r_{G,f}}=[0,0,1.3]^T$ km, 
$\mathbf{v_{G,f}}=[0,0,0]^T$ m/s, 
$\mathbf{a_{G,f}}=[0,0,3.2]^T$ m/s$^2$.}
\label{fig:fopdg_traj}
\end{figure}
We first compare the performance of the proposed analytical guidance with iterative fuel-optimal-powered descent guidance (FOPDG) as described in\cite{malyuta2022convex}. Fig \ref{fig:fopdg_traj} shows the time histories of position and velocity of the lander and Table \ref{tab:fuel_time} shows the performance comparison in terms of time of flight and fuel consumption.
\begin{table}[hbt!]
\caption{\label{tab:fuel_time} Performance Comparison}
\centering
\begin{tabular}{lcc}
\hline
\textbf{Guidance Algorithm} & \textbf{Fuel Consumed (kg)} & \textbf{Time of Flight (s)} \\
\hline
Proposed Approach & 154.5 & 164.4 \\
FOPDG & 152.1 & 154.8 \\
\hline
\end{tabular}
\end{table}
As discussed earlier, the proposed base guidance policy incurs a $2.7$ kg penalty, offline optimization identifies a fuel-optimal trajectory only within the class of functions parameterized by polynomial, but it offers a tractable analytical policy under state and action constraints.
\begin{figure}[t!] 
\centering 
    \begin{minipage}{.22\textwidth} 
    \centering 
    \includegraphics[width=1\textwidth]{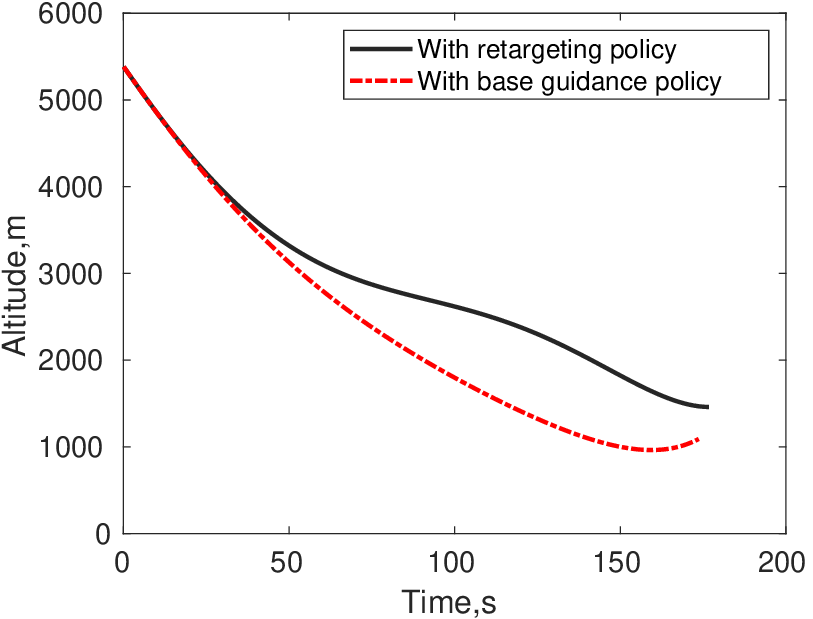} 
    \subcaption{Altitude}
    \label{fig:rtr_alt} 
    \end{minipage} 
    \begin{minipage}{.22\textwidth} 
    \centering 
    \includegraphics[width=1\textwidth]{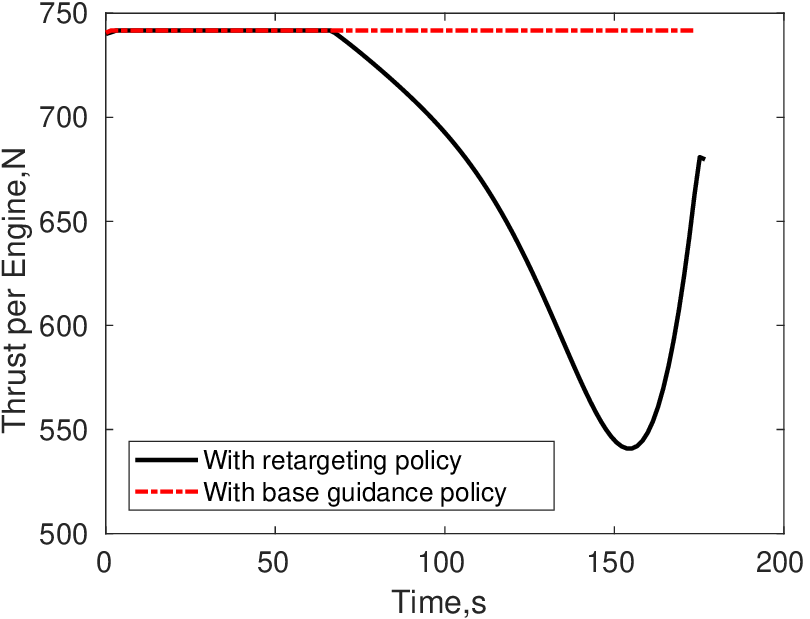} 
    \subcaption{Thrust Per Engine}
    \label{fig:rtr_thr} 
    \end{minipage} 
    \begin{minipage}{.22\textwidth} 
    \centering 
    \includegraphics[width=1\textwidth]{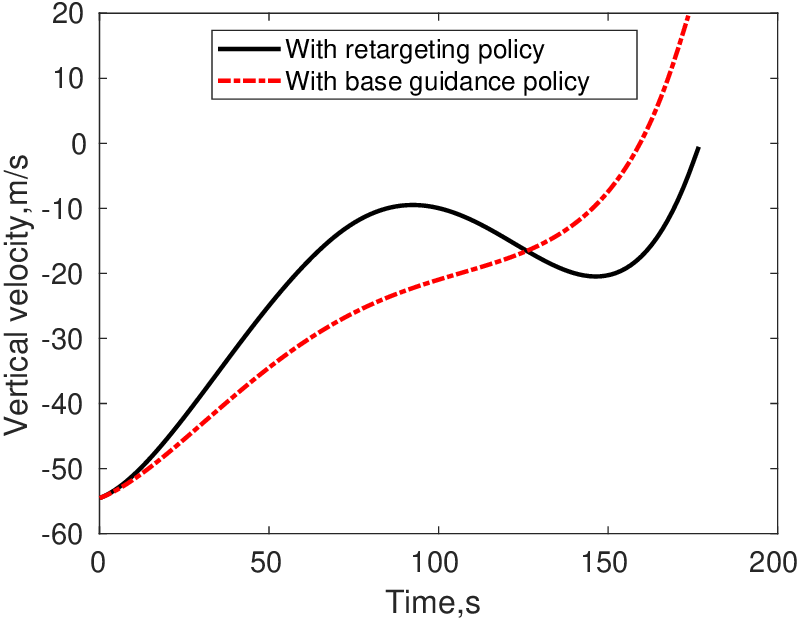} 
    \subcaption{Vertical Velocity}
    \label{fig:rtr_vv} 
    \end{minipage} 
    \begin{minipage}{.22\textwidth} 
    \centering 
    \includegraphics[width=1\textwidth]{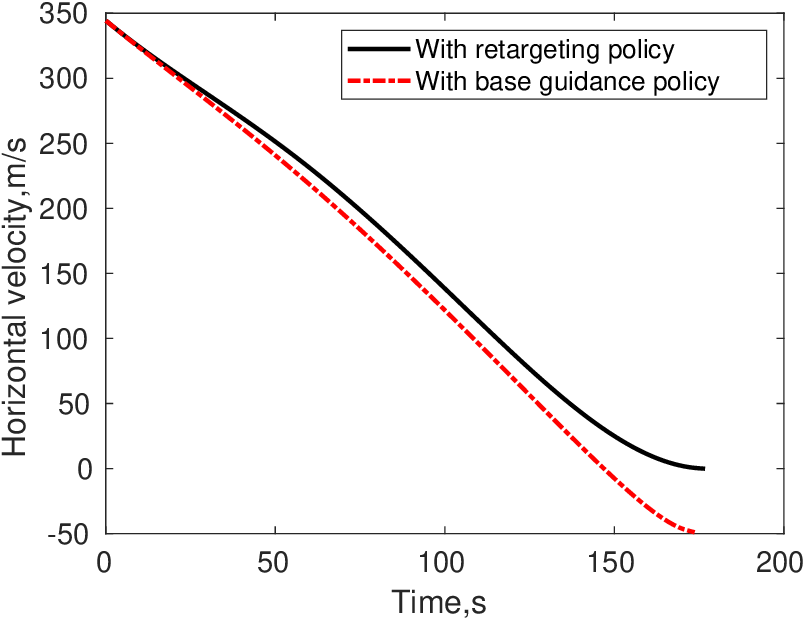} 
    \subcaption{Horizontal Velocity}
    \label{fig:rtr_hv} 
    \end{minipage} 
\caption{Comparison of trajectory and control trace between base policy and retargeting policy with simulation parameters: 
$\mathbf{r_{G,0}}=[-26.0,0,5.3]^T$ km, 
$\mathbf{v_{G,0}}=[344,0,-54]^T$ m/s, 
$\mathbf{a_{G,0}}=[-2.26,0.0,1.91]^T$ m/s$^2$; 
$\mathbf{r_{G,f}}=[0,0,1.3]^T$ km, 
$\mathbf{v_{G,f}}=[0,0,0]^T$ m/s, 
$\mathbf{a_{G,f}}=[0,0,3.2]^T$ m/s$^2$.}
\label{fig:rtr_traj}
\end{figure}
\begin{figure}[t!] 
\centering 
    \begin{minipage}{.22\textwidth} 
    \centering 
    \includegraphics[width=0.96\textwidth]{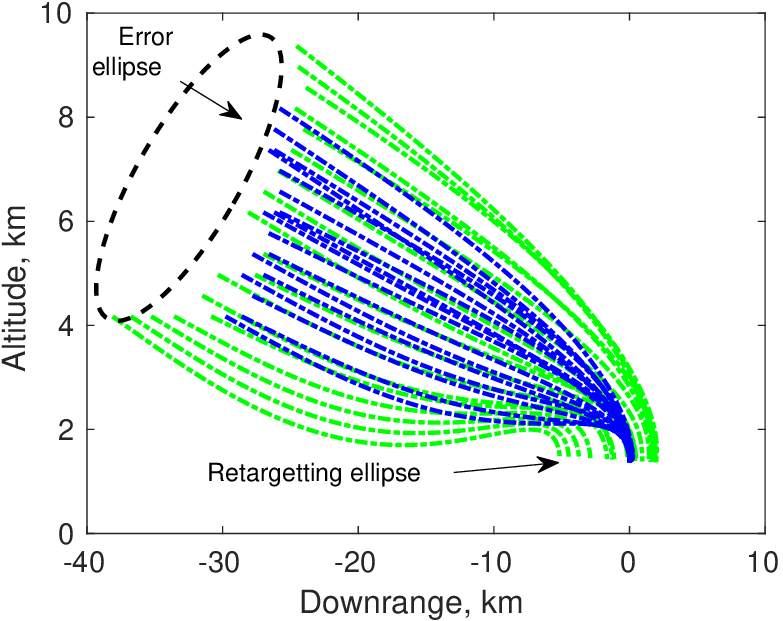} 
    \subcaption{Downrange vs Altitude}
    \label{fig:rtr_alt} 
    \end{minipage} 
    \begin{minipage}{.22\textwidth} 
    \centering 
    \includegraphics[width=1\textwidth]{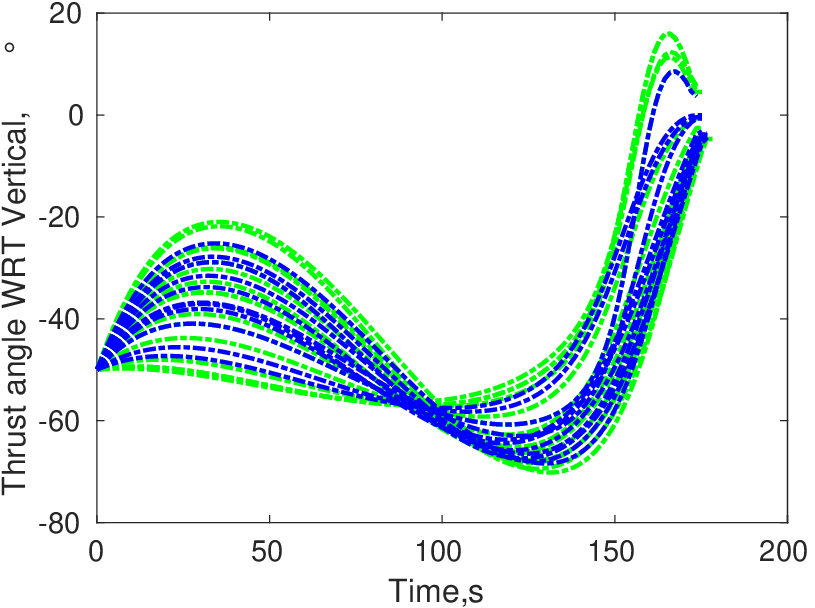} 
    \subcaption{Thrust Angle}
    \label{fig:rtr_thr} 
    \end{minipage} 
    \begin{minipage}{.22\textwidth} 
    \centering 
    \includegraphics[width=1\textwidth]{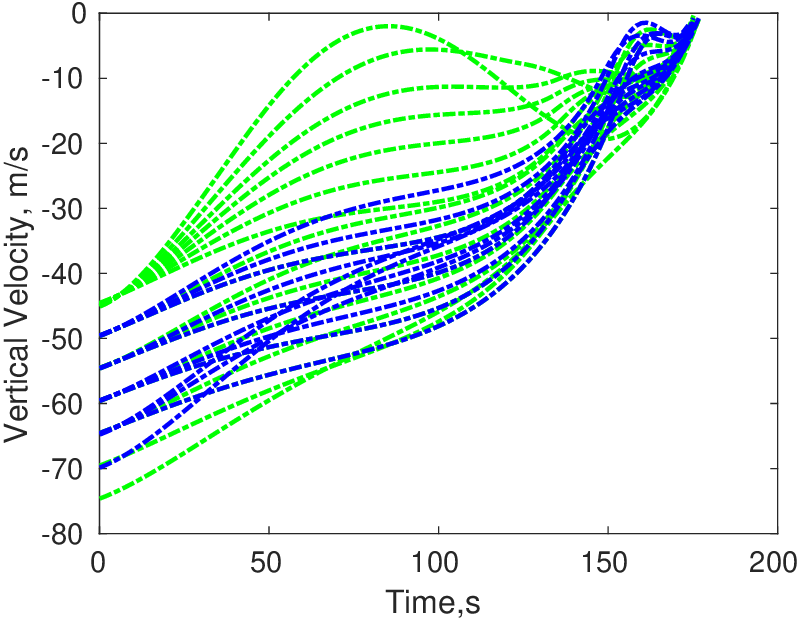} 
    \subcaption{Vertical Velocity}
    \label{fig:rtr_vv} 
    \end{minipage} 
    \begin{minipage}{.22\textwidth} 
    \centering 
    \includegraphics[width=1\textwidth]{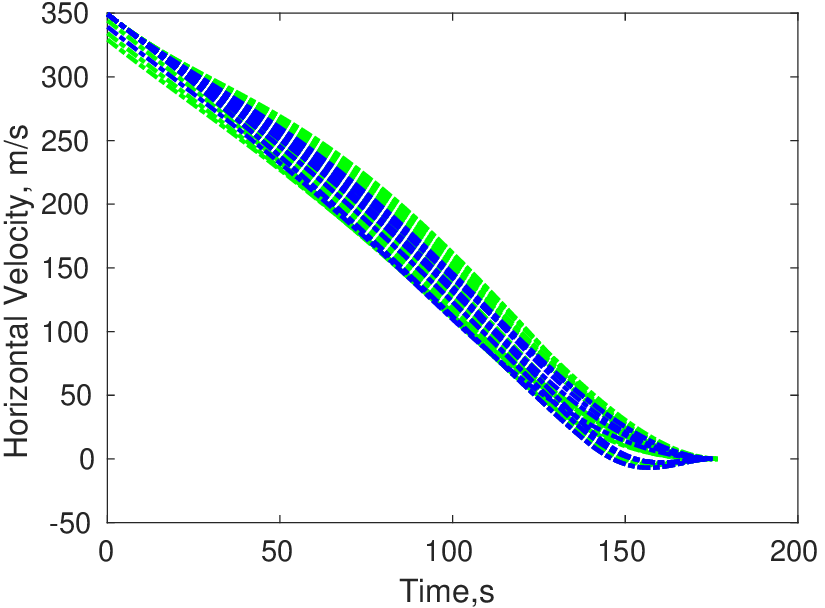} 
    \subcaption{Horizontal Velocity}
    \label{fig:rtr_hv} 
    \end{minipage} 
\caption{Trajectory trace for Monte-carlo simulation} \label{fig:mc_traj}
\end{figure}
To demonstrate the effectiveness of the real-time retargeting policy over the base policy, we consider a simulation with the initial guidance state is $2$ km ahead and $1.5$ km below the nominal state at the start of the fine braking phase. Fig.~\ref{fig:rtr_traj} compares the resulting trajectories under the base policy and the real-time retargeting policy. For this off-nominal condition, the base policy results in fully saturated thrust throughout the descent and fails to achieve the terminal constraints of $0$ m/s vertical velocity and $1.3$ km altitude. In contrast, the retargeting policy computes a target shift of $3.027$ km, enabling the lander to converge to the new target while satisfying all terminal requirements.

Monte Carlo simulations with randomly generated initial states are performed using the real-time retargeting policy. Fig.~\ref{fig:mc_traj} presents representative trajectory traces of range versus altitude, velocity, thrust, and thrust angle (cone angle) with respect to the vertical profiles. The blue curves correspond to cases where retargeting is not required, as the initial conditions lie within the controllable set. As the level of dispersion increases, pinpoint landing at the designated site becomes infeasible, which is detected by evaluating the initial condition against the controllability boundary. In such cases, retargeting to an alternate landing site is invoked, as illustrated by the green trajectories. 
\begin{figure}[t!] 
\centering 
    \begin{minipage}{.22\textwidth} 
    \centering 
    \includegraphics[width=0.94\textwidth]{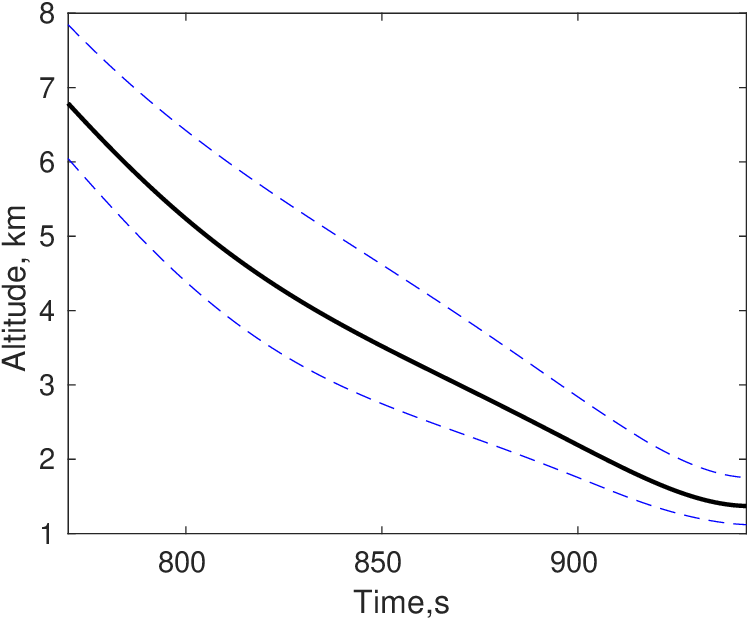} 
    \subcaption{Altitude}
    \label{fig:rtr_alt} 
    \end{minipage} 
    \begin{minipage}{.22\textwidth} 
    \centering 
    \includegraphics[width=1\textwidth]{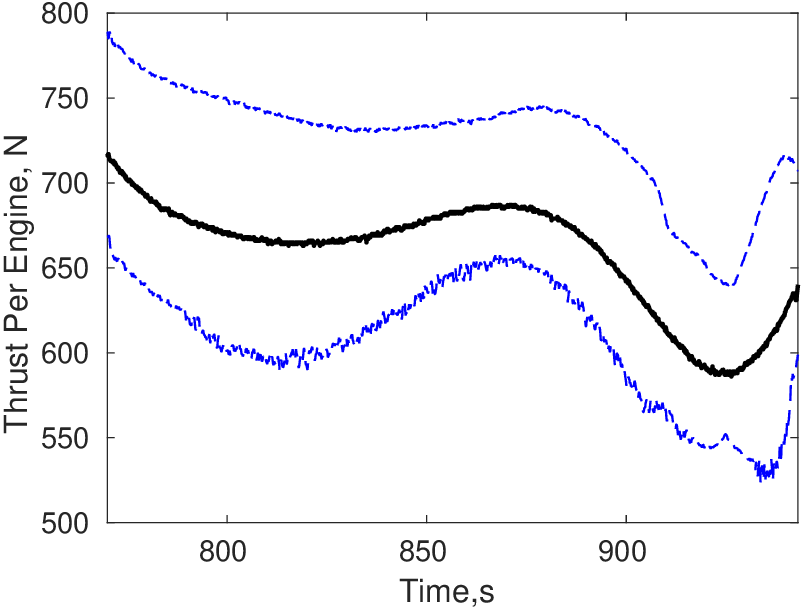} 
    \subcaption{Thrust Per Engine}
    \label{fig:rtr_thr} 
    \end{minipage} 
    \begin{minipage}{.22\textwidth} 
    \centering 
    \includegraphics[width=1\textwidth]{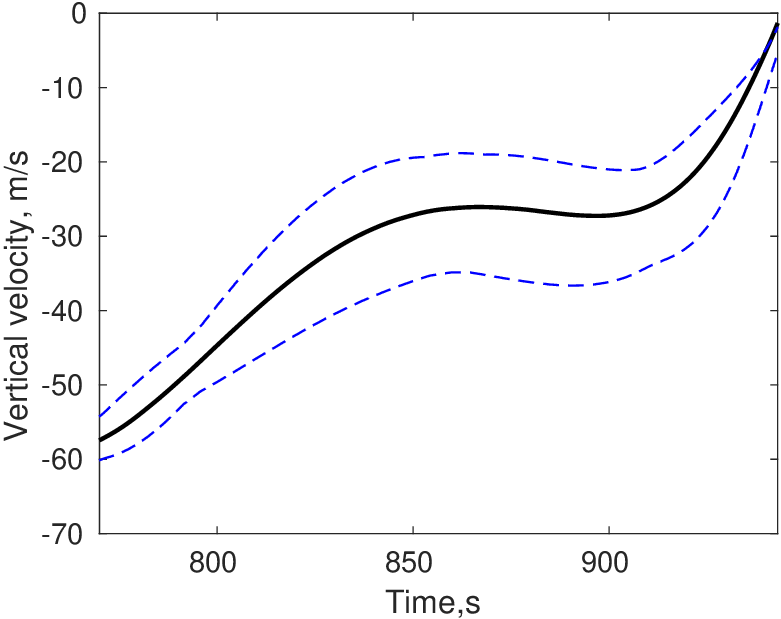} 
    \subcaption{Vertical Velocity}
    \label{fig:rtr_vv} 
    \end{minipage} 
    \begin{minipage}{.22\textwidth} 
    \centering 
    \includegraphics[width=1\textwidth]{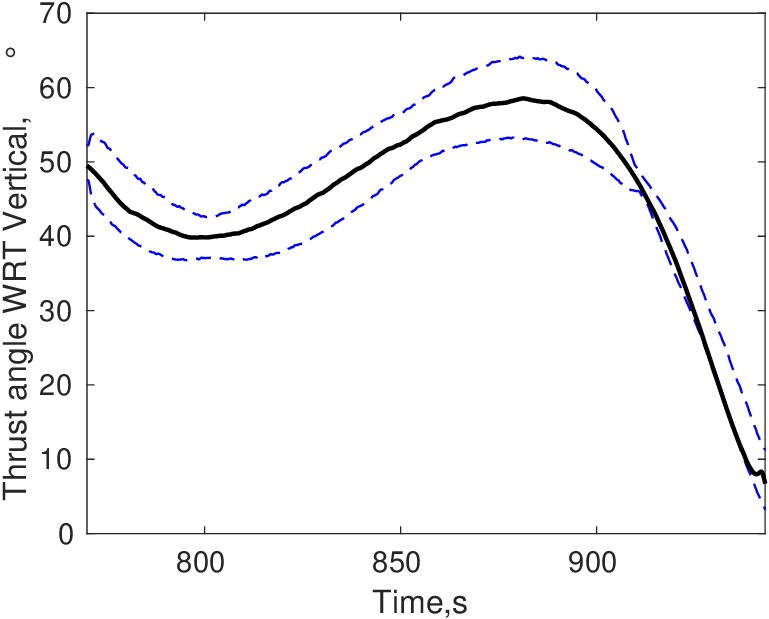} 
    \subcaption{Thrust Angle}
    \label{fig:rtr_hv} 
    \end{minipage} 
\caption{Flight results for the Fine braking phase} \label{fig:flight_traj}
\end{figure}
Finally, Fig. \ref{fig:flight_traj} shows the trajectory and control parameters from the flight performance of the fine braking phase. The blue traces denote the performance bounds predicted by Monte Carlo simulations of the proposed guidance approach, while the black traces represent the realized flight performance, which remains well within the anticipated bounds.

\section{CONCLUSIONS}
\label{sec:IV}
In this paper, we proposed a real-time retargeting guidance policy for lunar landing based on a convex representation of the controllability boundary. The base guidance policy solves an approximate fuel-optimal powered descent guidance problem via data-driven optimization within the class of accelerations parameterized by polynomials. Real-time retargeting policy computes its controllable set through bi-level convex optimization. This enables target shifts in real time when the designated landing site becomes infeasible, thereby enhancing robustness against state and control dispersions. Simulation results, including Monte Carlo analysis, demonstrate an increased safe landing ellipse, and thereby increasing the probability of mission success.

\section*{Acknowledgments}
We convey our sincere gratitude to the Indian Space Research Organization (ISRO) for encouraging and supporting this research. We wish to gratefully acknowledge the excellent review committee at ISRO for reviewing this work and providing valuable feedback. Authors convey sincere gratitude to the entire GNC and Chandrayaan-3 project team for their endless support during all major reviews.

\addtolength{\textheight}{-12cm}   % This command serves to balance the column lengths

%%%%%%%%%%%%%%%%%%%%%%%%%%%%%%%%%%%%%%%%%%%%%%%%%%%%%%%%%%%%%%%%%%%%%%%%%%%%%%%%

%%%%%%%%%%%%%%%%%%%%%%%%%%%%%%%%%%%%%%%%%%%%%%%%%%%%%%%%%%%%%%%%%%%%%%%%%%%%%%%%

%%%%%%%%%%%%%%%%%%%%%%%%%%%%%%%%%%%%%%%%%%%%%%%%%%%%%%%%%%%%%%%%%%%%%%%%%%%%%%%%
\bibliography{sample}

\end{document}